\documentclass[twocolumn,secnumarabic,amssymb, nobibnotes, aps, prd]{revtex4-1}

\usepackage{amsmath}
\usepackage{epsfig}
\usepackage{epic}
\usepackage{color}
\definecolor{bordo}{rgb}{0.8,0.3,0.3}
\definecolor{azul}{rgb}{0.1,0.2,0.7}
\usepackage[english]{babel}
\usepackage{color}
\usepackage{verbatim}
\definecolor{azul}{rgb}{0.1,0.2,0.6} 
\definecolor{verde}{rgb}{0.1,0.5,0.3}
\definecolor{bordo}{rgb}{0.9,0.3,0.3}
\usepackage{lipsum}
\usepackage{slashed}
\usepackage{tikz}
\usetikzlibrary{automata}
\usetikzlibrary{topaths}
\usepackage{pstricks}
\usetikzlibrary{automata,positioning,calc}
\usetikzlibrary{arrows}
\begin{document}

\title{\huge Equivalent Markov processes under gauge group}
\author{M. Caruso}%
\email{mcaruso@ugr.es}
\affiliation{Departamento de F\'isica Te\'orica y del Cosmos, Universidad de Granada, Campus de Fuentenueva, Granada (18071), Espa\~na.}

\author{C. Jarne}%

\email{cecilia.jarne@unq.edu.ar}
\affiliation{UNQ - Departamento de Ciencia y Tecnolog\'ia. \\IFIBA (UBA) Laboratorio de Sistemas Din\'amicos (CONICET).}


\begin{abstract}{We have studied Markov processes on denumerable state space and continuous time. We found that all these processes are connected via gauge transformations. We have used this result before as a method for resolution of equations, included the case where the sample space is time dependent in a previous work \textit{Phys. Rev. E} \textbf{90}, 022125 (2014). We found a general solution through a dilation of the state space, although the prior probability distribution of the states defined in this new space takes smaller values with respect to the one in the initial problem. The gauge (local) group of dilations modifies the distribution on the dilated space to restore the original process. In this work we show how Markov process in general could be linked via gauge (local) transformations and we present some illustrative examples for this results.}
\end{abstract}
\vspace{0.3cm}

\date{\today}

\maketitle

\section{Introduction}

Continuous-time Markov process are used to describe a variety of stochastic complex processes. They have been widely used in mathematical physics to describe the properties of important models in equilibrium and non-equilibrium, such as the Ising model \cite{Ising}.

Other important example of application is the use of continuous-time Markov chains in queueing theory \cite{Kleinrock}. 
Regarding biology, Markov chains are used to explain the properties of reaction networks, chemical system involving multiple reactions and chemical species \cite{Kurtz} and kinetics of linear arrays of enzymes \cite{Shapiro}.

In this work, we show a way to connect a given pair of Markov processes via gauge transformations. 
The link between different processes is a mathematical observation that enrich the description of the stochastic process. In addition, in some cases this observation could become an useful tool to study a particularly complex Markov problem using a simpler auxiliary Markov process and proposing an adequate transformation to link both of them.

This approach was heuristically explored on our recent work \cite{caraj} as an alternative method for the resolution of equations of Markov process on denumerable state spaces and continuous time. Nevertheless, in order to obtain a phenomenological or approximate solution to certain Markov process involves some knowledge of the system parameters and depends. This is precisely the case of \cite{caraj}. 

In the following we present a mathematical description of a general stochastic system. The structure of the paper is: in section \ref{sec-2}, a motivation of the problem in section \ref{sec-3}, the formal aspects of the equivalence of the process in section \ref{sec-4} and finally the conclusions and final comments in section \ref{sec-5}.


\section{Markovian process in a denumerable state space and continuous time} \label{sec-2}
	
We start by reviewing the basics aspects of this class of stochastic processes. Let's consider a stochastic system described by a Markov process with a random variable $x(t)$, which takes values from the state space at the \textit{instant} $t$
\begin{equation}\label{genstate}
\mathcal{S}=\{x_n: n\in \pmb{\ell}\},
\end{equation}
where $t$ represent a \textit{time variable}, some parameter used to describe the evolution of the process, that takes values from a set $\mathcal{T}\subseteq\mathbb{R} $, $\pmb{\ell}$ is the countable set of \textit{labels} for the states, such that $\pmb{\ell}\subseteq\mathbb{Z}_0^+$.

We defined the conditional probability to find the system in the state $x_l$, at the instant $t$, given that at instant $s$ was in the state $x_k$, denoted by
\begin{equation}
\mathcal{P}_{lk}(t,s)=\mathbb{P}\pmb{(}x(t)=x_l|x(s)=x_k\pmb{)}.
\end{equation} 

We understand this conditional probability as a transition element between the states $x_k\longmapsto x_l$ and with a temporal evolution $s\longmapsto t$. These conditional probabilities describes the time evolution of the stochastic system, in the sense that they allow us to connect any two ordered pairs  $(x_k,s)$, $(x_l,t)$.

The time evolution of a Markov process, is determined by the knowledge of a prior probability distribution for each $t$, denoted by 
\begin{equation}
p_n(t)=\mathbb{P}\pmb{(}x(t)=x_n\pmb{)},
\end{equation} 
for all $(t,n)\in \mathcal{T}\times\pmb{\ell}$.

An equivalent way to describe this process is through an initial value $p_n(0)$ and a conditional probabilities $\mathcal{P}_{nm}(t,s)$, which represents the transition matrix elements of the states $x_m\longmapsto x_n$. For each $t$ the events are mutually exclusive, then 
\begin{equation}
p_n(t)=\sum_{m\in \pmb{\ell}}\mathcal{P}_{nm}(t,s)\:p_m(s).
\end{equation}
	
Consequently at the time $t+\epsilon$ the probability to find the system in $x_n$, is given by 	the transition from $x_{m}$ at time $t$, in this way
\begin{equation}\label{super}
p_n(t+\epsilon)=\sum_{m\in \pmb{\ell}}\mathcal{P}_{nm}(t+\epsilon,t)\:p_m(t).
\end{equation}

After some elementary operations, we get: 
\begin{equation}\label{evo}
d_tp_n(t)=\sum_{m\in \pmb{\ell}}\mathtt{Q}_{nm}(t) \:p_m(t),
\end{equation}
where $d_t$ is a notation of total time derivative, and $\mathtt{Q}_{nm}(t)$ is given by
\begin{equation}
\mathtt{Q}_{nm}(t)=\partial_t \mathcal{P}_{nm}(t,s)|_{s=t}.
\end{equation}
where $\mathtt{Q}_{nm}(t)$ is called the \textit{infinitesimal generator}.

The equation \eqref{evo} is named the Kolmogorov equation, the foundational work \cite{Kolmogorov}. Another authors are referred later to \eqref{evo} as the forward Kolmogorov equations \cite{Feller}.
 
We define $\pmb\varphi(t)$ as an $|\mathcal{S}|$-tuple of the probability distribution as
$\pmb\varphi(t)=(\:p_0(t),p_1(t),\cdots\:)^\intercal$. And also we used a notation for the cardinal number of a set $\pmb{S}$ is given by $|\pmb{S}|$ and $^\intercal$ represents the transposition operation.

The evolution equation for the process can be expressed in a \textit{matrix form} as \cite{Masaaki}
\begin{equation}\label{evolu}
d_t\pmb\varphi(t)=\pmb{\mathtt{Q}}(t)\:\pmb\varphi(t).
\end{equation}

In this way we have a mathematical description of a Markov process in terms of a set of prior probabilities $\{p_0(t),p_1(t),\cdots\}$ and an infinitesimal generator $\pmb{\mathtt{Q}}(t)$. 
\section{Motivation of the problem} \label{sec-3}

In this section we present a motivational example to show how a given pair of Markov processes could be linked via gauge transformations. Let's consider two particular stochastic processes, with their respective infinitesimal generators 
\begin{align}
\pmb{\mathtt{Q}}_+=\left(\begin{matrix}\label{Q+-}
-\nu & 0  \\ 
 \;\;\nu & 0 
\end{matrix}\right),\;
\pmb{\mathtt{Q}}_-=\left(\begin{matrix}
0&\;\;\mu\\ 
0&-\mu  
\end{matrix}\right).
\end{align}

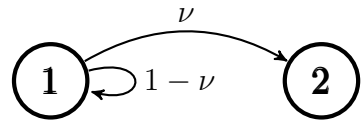
\begin{figure}[h!]
\begin{center}
\begin{tikzpicture}[->, >=stealth', auto, thick, node distance=3cm]
\tikzstyle{every state}=[fill=white,draw=black,ultra thick,text=black,scale=1.2]
\node[state]    (A) {\large ${\pmb 1}$};
\node[state]    (D)[right of=A]   {\large $\pmb{2}$};
\path
(A) edge[loop right]     node{\large$1-\nu$}(A)
(A) edge[bend left,above]node{\large$\nu$}  (D);
\end{tikzpicture}
\end{center}
\caption{\footnotesize{Diagram associated to the finite pure birth process with the infinitesimal generator $\pmb{\mathtt{Q}}_+$.}}\label{diag+}
\end{figure}

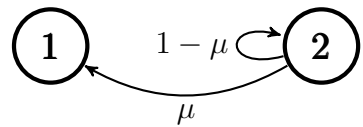
\begin{figure}[h!]
\begin{center}
\begin{tikzpicture}[->, >=stealth', auto, thick, node distance=3cm]
\tikzstyle{every state}=[fill=white,draw=black,ultra thick,text=black,scale=1.2]
\node[state]    (A) {\large ${\pmb 1}$};
\node[state]    (D)[right of=A]   {\large $\pmb 2$};
\path
(D) edge[loop left]     node{\large$1-\mu$}(D)
(D) edge[bend left,below]node{\large$\mu$}  (A);
\end{tikzpicture}
\end{center}
\caption{\footnotesize{Diagram associated to the finite pure death process with the infinitesimal generator $\pmb{\mathtt{Q}}_-$.}}\label{diag-}
\end{figure}

This matrices $\pmb{\mathtt{Q}}_+$ and $\pmb{\mathtt{Q}}_-$ corresponds to a \textit{pure birth process} and a \textit{pure death process}, whereas $\nu$ and $\mu$ are the birth and death rates, respectively. Note that $\pmb{\mathtt{Q}}_++\,\pmb{\mathtt{Q}}_-$, from \eqref{Q+-}, is equal to another infinitesimal generator which corresponds to a finite $2-$state birth-death process. Also, to be more explicit, we can represent each of these processes through the diagrams in Fig. \ref{diag+} and \ref{diag-}.

\vspace{0.5cm}

The differential equation from \eqref{evolu} applied for each of these processes are summarized by:
\begin{align}\label{evolu+-}
d_t\pmb\varphi_+(t)&=\pmb{\mathtt{Q}}_+(t)\:\pmb\varphi_+(t),\nonumber \\
&\\
d_t\pmb\varphi_-(t)&=\pmb{\mathtt{Q}}_-(t)\:\pmb\varphi_-(t).\nonumber 
\end{align}

We demonstrate that exist a $2\times 2$ matrix $\pmb{\lambda}$ that connect the solutions $\pmb\varphi_+$ and $\pmb\varphi_-$ in a following way:
\begin{equation}\label{demostrar}
\pmb\varphi_+=\pmb\lambda\pmb\varphi_- .
\end{equation}
First of all only for the particular case $\nu=\mu$ we will have a constant matrix $\pmb{\lambda}$
\begin{align}\label{lambdareflection}
\pmb{\lambda}&=\left(\begin{matrix}
0&1\\ 
1&0 
\end{matrix}\right),
\end{align}
which corresponds to an interchange of the states \tikz[baseline=(char.base)]{\node[shape=circle,draw,thick,inner sep=2pt](char){\textbf{1}};}$\longleftrightarrow$\tikz[baseline=(char.base)]{\node[shape=circle,draw,thick,inner sep=2pt](char){\textbf{2}};}. In other words, \eqref{lambdareflection} corresponds to a reflection that interchange the Fig. \ref{diag+} by the Fig. \ref{diag-} and viceversa. Explicitely for the case $\nu=\mu$, using \eqref{lambdareflection}, we have 
\begin{align}\label{a}
\pmb{\mathtt{Q}}_+&=\pmb{\lambda}\;\pmb{\mathtt{Q}}_-\;\pmb{\lambda}^{-1}.
\end{align}
The last equation together with \eqref{evolu+-} involve \eqref{demostrar}. The matrix $\pmb{\lambda}$ from \eqref{lambdareflection} is a time independent change of coordinates between equations \eqref{evolu+-} for the case $\nu=\mu$.

We have noticed that $\pmb\lambda$ is not a constant matrix, for the case $\nu\neq\mu$. For the present  case we obtain the solutions with a non trivial initial conditions are
\begin{align}
\pmb{\varphi}_+&=\begin{pmatrix}\label{phi+,phi-}
e^{-\nu\,t} \\ 
\\
1-e^{-\nu\,t} 
\end{pmatrix},\quad 
\pmb{\varphi}_-=\begin{pmatrix}
1-e^{-\mu\,t} \\ 
\\
e^{-\mu\,t} 
\end{pmatrix}.
\end{align}
Using the explicit solutions \eqref{phi+,phi-}, the proof that the equation \eqref{demostrar} is true is straightforward, since there is a matrix  
\begin{align}\label{lambda2x2}
\pmb{\lambda}&=\left(\begin{matrix}
0& & e^{(\mu-\nu)\,t}\\ 
 & & \\
1& & 1-e^{(\mu-\nu)\,t} 
\end{matrix}\right),
\end{align}
such that the equation \eqref{demostrar} is true.

This example shows, in pedagogical way, that it is possible to write the solution of a stochastic process starting from another process. The bridge between $\pmb\varphi_+$ and $\pmb\varphi_-$ is built through a local transformation $\pmb\lambda$. Also the correspond link between $\pmb{\mathtt{Q}}_+$ and $\pmb{\mathtt{Q}}_-$ through a local transformation $\pmb\lambda$ is given by
\begin{equation}\label{b} 
\pmb{\mathtt{Q}}_+=\pmb{\lambda} \pmb{\mathtt{Q}}_- \pmb{\lambda}^{-1}+  d_t\pmb{\lambda} \pmb{\lambda}^{-1},
\end{equation}
for all $\nu$ and $\mu$. The expression \eqref{b} is almost equal to \eqref{a} but with an added term $d_t\pmb{\lambda} \pmb{\lambda}^{-1}$.

We will see that the group of this kind of transformations is structured as a gauge group. 
We can formalize and generalize this idea in the following sections and we prove that is possible to connect any pair of infinitesimal generator $(\pmb{\mathtt{Q}},\pmb{\mathtt{Q}}')$ and any pair of prior distributions of probability $(\pmb{\varphi},\pmb{\varphi}')$, associated to these infinitesimal generators, in a similar way that \eqref{b} and \eqref{demostrar}, respectively
\begin{align}
&\pmb{\mathtt{Q}}'=\pmb{\lambda} \pmb{\mathtt{Q}} \pmb{\lambda}^{-1}+  d_t\pmb{\lambda} \pmb{\lambda}^{-1},\nonumber \\
&\\
&\pmb{\varphi}'=\pmb{\lambda}\pmb{\varphi}.\nonumber	
\end{align}
\section{Formal aspects of equivalent Markov Processes }\label{sec-4}

We considered a map $\pmb{\Gamma_\lambda}$,  given a non singular matrix $\pmb{\lambda}$, which transforms a matrix $\pmb{\mathtt{Q}}$ as
\begin{equation}\label{map}
\pmb{\Gamma}_{\pmb{\lambda}}(\pmb{\mathtt{Q}}) = \pmb{\lambda} \pmb{\mathtt{Q}} \pmb{\lambda}^{-1}+  d_t\pmb{\lambda} \pmb{\lambda}^{-1},
\end{equation}
where $\pmb{\lambda}$, $\pmb{\mathtt{Q}}\in \mathbb{R}^{|\pmb{\ell}|\times|\pmb{\ell}|}$ are $t-$dependent differentiable matrices. Thereby, $\pmb{\lambda}$ is a local transformation and we will prove that $\pmb{\Gamma_\lambda}$ form a group of local (gauge) transformations. In particular $\pmb{\Gamma_\lambda}\in \mathbb{R}^{|\pmb{\ell}|\times|\pmb{\ell}|}$.

In addition, in this section we study the possibility that for all pair of matrices $\pmb{\mathtt{Q}}$ and $\pmb{\mathtt{Q}}'$, $t-$dependent and differentiable, there is a non singular matrix $\pmb{\lambda}$, $t-$dependent and differentiable, that connect $\pmb{\mathtt{Q}}$ and $\pmb{\mathtt{Q}}'$ as 
\begin{equation}\label{connect}
\pmb{\mathtt{Q}}'=\pmb\Gamma_{\pmb{\lambda}} (\pmb{\mathtt{Q}}).
\end{equation}

If we composed two transformation $\pmb{\Gamma_{\lambda}}\circ\pmb{\Gamma}_{\pmb{\lambda}'}$ with
$\pmb{\lambda}$ and $\pmb{\lambda}'$ are non singular, we see that 
\begin{align}\label{composegamma}
\pmb{\Gamma_{\lambda}}\circ\pmb{\Gamma}_{\pmb{\lambda}'}(\pmb{\mathtt{Q}})
& = \pmb{\Gamma}_{\pmb{\lambda}\pmb{\lambda}'}(\pmb{\mathtt{Q}}).
\end{align}

From \eqref{composegamma} we see that if 
\begin{equation}\label{commutator}
[\pmb{\lambda,\lambda}']=\pmb{0}\Longrightarrow\pmb{\Gamma_{\lambda}}\circ\pmb{\Gamma}_{\pmb{\lambda}'}(\pmb{\mathtt{Q}})=\pmb{\Gamma}_{\pmb{\lambda}'}\circ\pmb{\Gamma}_{\pmb{\lambda}}(\pmb{\mathtt{Q}}).
\end{equation}

Using this properties of composition \eqref{composegamma} and \eqref{commutator},  we give an expression for the inverse map  $\pmb{\Gamma_{\lambda}}^{-1}$. First of all, we have trivially
\begin{equation}
\pmb{\Gamma_{1}}(\pmb{\mathtt{Q}})=\pmb{\mathtt{Q}},
\end{equation}
where $\pmb{1}$ is the identity matrix.  If we consider the composed transform $\pmb{\lambda}''=\pmb{\lambda\lambda}'$ such that $\pmb{\lambda\lambda}'=\pmb{1}=\pmb{\lambda}'\pmb{\lambda}$ then from \eqref{commutator} we have 
\begin{equation}\label{unique inv} 
\pmb{\Gamma_{\lambda}}\circ\,\pmb{\Gamma}_{\pmb{\lambda}'}=\pmb{1}=\pmb{\Gamma}_{\pmb{\lambda}'}\circ\,\pmb{\Gamma}_{\pmb{\lambda}}.
\end{equation}
Finally from \eqref{unique inv} the inverse of $\pmb{\Gamma}_{\pmb{\lambda}}$ is \textit{unique} and given by 
\begin{align}\label{Gammainv}
\pmb{\Gamma_{\lambda}}^{-1}=\pmb{\Gamma}_{\pmb{\lambda}^{-1}},
\end{align}
for more details of the properties of composition \eqref{composegamma} and inverse transformation \eqref{Gammainv} see Appendix \textbf{A1}. 

We will demonstrate that for any pair of $t-$dependent differentiable matrices $\pmb{\mathtt{Q}}$ and $\pmb{\mathtt{Q}}'$ both of $|\pmb{\ell}|\times|\pmb{\ell}|$ there exist a non-singular $t-$dependent differentiable matrices $\pmb{\lambda}$ of $|\pmb{\ell}|\times|\pmb{\ell}|$ that connect them. For that we can define the following \textit{equivalence relation}:
\begin{align}\label{equivrel}
\pmb{\mathtt{Q}}'\sim \pmb{\mathtt{Q}}\Longleftrightarrow \exists \pmb{\lambda}: \pmb{\mathtt{Q}}'=\pmb{\Gamma}_{\pmb{\lambda}}(\pmb{\mathtt{Q}}),
\end{align}
more details that $\sim$ is a well defined \textit{equivalence relation} see Appendix \textbf{A1}.  
From the equivalence relation \eqref{equivrel} then  $\pmb{\lambda}$ satisfies the differential equation:
\begin{align}\label{lambdaeq}
d_t\pmb{\lambda}=\pmb{\mathtt{Q}}'\pmb{\lambda}-\pmb{\lambda \mathtt{Q}}.
\end{align}


First of all, the solution of \eqref{lambdaeq} exist for the trivial cases $\pmb{\mathtt{Q}}=\pmb{0}$ and $\pmb{\mathtt{Q}}'=\pmb{0}$, i.e. we denoted by $\pmb{\lambda}_1$ and $\pmb{\lambda}_2$ the respective solutions for each case 
\begin{align}
d_t\pmb{\lambda}_1&=\pmb{\mathtt{Q}}'\pmb{\lambda}_1,\label{l1}\\
&\nonumber \\
d_t\pmb{\lambda}_2&=-\pmb{\lambda}_2\pmb{\mathtt{Q}}.\label{l2}
\end{align}
We can obtain $(\pmb{\lambda}_1,\pmb{\lambda}_2)$ as an iterative non singular solutions, for more details of this solution see Appendix \textbf{A2}. The existence of solutions for \eqref{l1} and \eqref{l2} implies that $\pmb{\lambda}_1$ and $\pmb{\lambda}_2$ connects $\pmb{\mathtt{Q}}' \sim \pmb{0}$ and $\pmb{0} \sim \pmb{\mathtt{Q}}$, respectively. This implication is true from the definition of the equivalence relation. From the existence of solutions for \eqref{l1} and \eqref{l2} then we have 
\begin{align}
\exists \pmb{\lambda}_1: \pmb{\mathtt{Q}}'=\pmb{\Gamma}_{\pmb{\lambda}_1}(\pmb{0})
&\Longleftrightarrow\pmb{\mathtt{Q}}'\sim \pmb{0},\label{transitive1}\\
&\nonumber \\
\exists \pmb{\lambda}_2: \pmb{0}=\pmb{\Gamma}_{\pmb{\lambda}_2}(\pmb{\mathtt{Q}})
&\Longleftrightarrow\pmb{0}\sim\pmb{\mathtt{Q}},\label{transitive2}
\end{align} 
and from transitivity of the equivalence relation \eqref{equivrel} we have $\pmb{\mathtt{Q}}'\sim \pmb{\mathtt{Q}}$, this means that there is a given $\pmb{\lambda}$ that $\pmb{\mathtt{Q}}'=\pmb{\Gamma}_{\pmb{\lambda}}(\pmb{\mathtt{Q}})$. 

We express the solution $\pmb{\lambda}$ as a function of the solutions of \eqref{l1} and \eqref{l2}, $(\pmb{\lambda}_1,\pmb{\lambda}_2)$, respectively. We say that a solution $\pmb{\lambda}$ built in this way is a \textit{transitive solution}, or \textit{composite solution}, the name will be clear in the construction procedure of the solution $\pmb{\lambda}$. From \eqref{transitive1} and \eqref{transitive2} we see that the \textit{transitivity solution} is constructed from the composition of transformations $\pmb{\mathtt{Q}}'=\pmb{\Gamma}_{\pmb{\lambda}_1}(\pmb{0})$ and $\pmb{0}=\pmb{\Gamma}_{\pmb{\lambda}_2}(\pmb{\mathtt{Q}})$ as follows
\begin{align}\label{comp}
\pmb{\mathtt{Q}}'=\pmb{\Gamma}_{\pmb{\lambda}_1}\pmb{(}\pmb{\Gamma}_{\pmb{\lambda}_2}(\pmb{\mathtt{Q}})\pmb{)}
\end{align}
from the composition rule \eqref{composegamma} applied to \eqref{comp}
\begin{align}
\pmb{\mathtt{Q}}'=\pmb{\Gamma}_{\pmb{\lambda}_1\pmb{\lambda}_2}(\pmb{\mathtt{Q}})
\end{align}
where the \textit{transitive solution} is given by
\begin{align}\label{lambdacomp}
\pmb{\lambda}=\pmb{\lambda}_1\pmb{\lambda}_2.
\end{align}

We have demonstrated that for any pair of matrices $(\pmb{\mathtt{Q}},\pmb{\mathtt{Q}}')$, $t-$dependent and differentiable, there is a non singular matrix $\pmb{\lambda}$, $t-$dependent and differentiable, that connect $\pmb{\mathtt{Q}}$ and 
$\pmb{\mathtt{Q}}'$ through the map $\pmb{\Gamma_\lambda}$, given by the expression \eqref{map}
\begin{equation}\label{Qequiv}
\pmb{\mathtt{Q}}'=\pmb{\Gamma_\lambda}(\pmb{\mathtt{Q}}).
\end{equation}

Suppose now that $\pmb{\mathtt{Q}}$ and $\pmb{\mathtt{Q}}'$ are the infinitesimal generators, $t-$dependent and differentiable, of the following differential equations
\begin{align}\label{equations}
d_t\pmb{\varphi}&=\pmb{\mathtt{Q}}\;\pmb{\varphi},\nonumber\\
&\\
d_t\pmb{\varphi}'&=\pmb{\mathtt{Q}}'\pmb{\varphi}',\nonumber 
\end{align}
finally from \eqref{Qequiv} and \eqref{equations} we have 
\begin{align}\label{conlc}
\pmb{\varphi}'=\pmb{\lambda}\pmb{\varphi}.
\end{align}

We found that for any pair infinitesimal generators, $t-$dependent and differentiable, $(\pmb{\mathtt{Q}},\pmb{\mathtt{Q}}')$ associated to \eqref{equations}, exist another $t-$dependent and differentiable matrix $\pmb \lambda$ that connect the distribution of probability $\pmb{\varphi}$ and $\pmb{\varphi}'$.

Until now we have considered the equivalence of Markov processes of the same dimension, i.e. the state spaces of every couple of processes have the same cardinality. We will go one step further, we will prove the equivalence of all continuous-time Markov processes on a denumerable state space. 

Without loss of generality we define $(\pmb{\mathtt{Q}},\mathcal{S})$ and $(\pmb{\mathtt{Q}}',\mathcal{S}')$ as the respective infinitesimal generators and state spaces, such that $n=|\mathcal{S}|<|\mathcal{S}'|=n'$. We can construct another process associated with $\pmb{\mathtt{Q}}$, such that the infinitesimal generator, $\pmb{\mathbf{Q}}$, is given by
\begin{align}\label{gen inf artif}
\mathbf{Q}_{ij}=\Bigg{\{} \begin{matrix}
\mathtt{Q}_{ij}, &  &\forall i,j\in [1,n]\subseteq \mathbb{N}\\
0, &  &\forall i,j   \in [n+1,n']\subseteq \mathbb{N}
\end{matrix}
\end{align}
or in block form
\begin{align}\label{gen inf artif2}
\pmb{\mathbf{Q}}=\left(\begin{matrix}
\pmb{\mathtt{Q}} &&   0    & &  \hdots\\
0& & 0 & &\hdots\\
\vdots & & \vdots &&\ddots
\end{matrix}\right).
\end{align}
The matrix $\pmb{\mathbf{Q}}$ corresponds to a new process on a state space $\mathbf{S}$ which have the same cardinality of $\mathcal{S}'$. We have completed the process on $\mathcal{S}$ with a number of \textit{redundant} states, such that the resulting state space $\mathbf{S}$ satisfy $|\mathbf{S}|=n'=|\mathcal{S}'|$. For illustrative purposes the Fig. \ref{complete} shows the composed state space $\mathbf{S}$ from $\mathcal{S}$ and a set of isolated and absorbing states $\{$\tikz[baseline=(char.base)]{\node[shape=circle,draw,thick,inner sep=0.9pt](char){\footnotesize $k$};} : $k\in[n+1,n']\subseteq \mathbb{N}\}$:
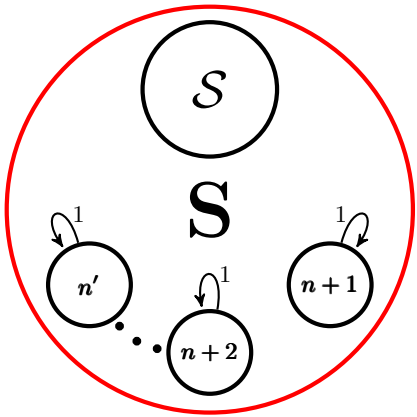
\begin{figure}[h!]
\begin{center}
\begin{tikzpicture}[->, >=stealth', auto, thick, node distance=1.5cm]
\draw[red,ultra thick] (0,0) circle (2.7cm);
\node[xshift=0cm,yshift=0cm][fill=white, very thick,text=black,scale=2.5](S) {\large $\mathbf{S}$};
\node[state,xshift=0cm,yshift=1.6cm][fill=white,draw=black, ultra thick,text=black,scale=2.2](A)  {$\mathcal{S}$};
\tikzstyle{every state}=[fill=white,draw=black,ultra thick,text=black,scale=1, minimum size=1.1cm]
\node[state]    (B)[xshift=1.6cm,yshift=-1cm]   {$\pmb{n+1}$};
\node[state]    (C)[xshift=0cm,yshift=-1.9cm]   {$\pmb{n+2}$};
\node[state]    (D)[xshift=-1.6cm,yshift=-1cm]  {$\pmb{n'}$};
\draw[black,ultra thick] (-0.7cm,-1.85cm)   circle (0.03cm);
\draw[black,ultra thick] (-0.97cm,-1.75cm) circle (0.03cm);
\draw[black,ultra thick] (-1.2cm,-1.55cm)   circle (0.03cm);
\path
(B) edge[out=75,in=55,loop, left]node{$1\;$}(B)
(C) edge[in=100,out=80,loop, right] node{1}(C)
(D) edge[in=90+35,out=90+15,loop, right	] node{$\;1$}(D);

\end{tikzpicture}
\end{center}
\caption{(Color online) \footnotesize{This diagram shows the composed process on the state space $\mathbf{S}\approx \,\mathcal{S}'$, in the sense that $|\mathbf{S}|=|\mathcal{S}'|$, with the infinitesimal generator $\pmb{\mathbf{Q}}$.}}\label{complete}
\end{figure}

All the states of $\{$\tikz[baseline=(char.base)]{\node[shape=circle,draw,thick,inner sep=0.9pt](char){\footnotesize $k$};} : $k\in[n+1,n']\subseteq \mathbb{N}\}$ are isolated or mutually disconnected and also from each state of $\mathcal{S}$; they are  all absorbing states. If the process start in some \tikz[baseline=(char.base)]{\node[shape=circle,draw,thick,inner sep=0.9pt](char){\footnotesize $k$};} of this redundant set, it stays there forever. 

A final comment is related to the case where the state space is time dependent Markov process. This is in the sense that its cardinality is a function of time $n_t$, meaning that for a given generation time $t$ (this is $t\in\mathbb{N}$) the state space $\mathcal{S}_t$: $|\mathcal{S}_t|=n_t$. Explicitly we have 
\begin{equation}
\mathcal{S}_t=\{x_1,x_2,\cdots,x_{n_t}\},
\end{equation}
for a given $t$.
If $\pmb{\mathtt{Q}}_t$ is the infinitesimal generator for each $t$, we can construct another process associated with $\pmb{\mathtt{Q}}_t$, such that the infinitesimal generator, $\pmb{\mathbf{Q}}$, is constructed in a similar way that \eqref{gen inf artif}
\begin{align}\label{gen inf artif3}
\mathbf{Q}_{ij}=\Bigg{\{} \begin{matrix}
(\mathtt{Q}_t)_{ij}, &  &\forall i,j\in [1,n_t]\subseteq \mathbb{N}\\
0, &  &\forall i,j   \in [n_t+1,N]\subseteq \mathbb{N}
\end{matrix}
\end{align}
for sufficiently big number $N\in\mathbb{Z}_0^+$. In other terms we can write
\begin{align}
\mathbf{Q}_{ij}=(\mathtt{Q}_t)_{ij}\;u(n_t-i)u(n_t-j),
\end{align}
where $u(x)$ is a Heaviside step function
\begin{align}\label{Heaviside}
u(x):=\Bigg{\{} \begin{matrix}
0:&x<0\\
1:&x\geq0
\end{matrix}\;.	
\end{align}
In a matrix form we express the dilution of $\mathcal{S}_t$ inside $\mathbf{S}=\{x_1,x_2,\cdots,x_N\}$, for finite value of $N$ or $N \rightarrow\infty$:
\begin{align}\label{gen inf artif4}
\pmb{\mathbf{Q}}=\left(\begin{matrix}
\pmb{\mathtt{Q}}_t &&   0    & &  \hdots\\
0& & 0 & &\hdots\\
\vdots & & \vdots &&\ddots
\end{matrix}\right).
\end{align}
The matrix $\pmb{\mathbf{Q}}$ corresponds to a new process on a state space $\mathbf{S}$, which it is equivalent to any other.

\section{Conclusion and final observations} \label{sec-5}

The aim of present work is to demonstrate that there is a way to modify the solution for a simple or known process, which is represented by the infinitesimal generator $\pmb{\mathtt{Q}}$, in order to get another process partially known, or at least with a very profound difficulty to be resolved, and represented by $\pmb{\mathtt{Q}}'$. 

We have shown how for a given pair of Markov processes $(\pmb{\mathtt{Q}},\mathcal{S})$ and $(\pmb{\mathtt{Q}}',\mathcal{S}')$ they could be linked via gauge (local) transformations $\pmb{\lambda}$, that allow us obtain $\pmb{\mathtt{Q}}'$ from $\pmb{\mathtt{Q}}$ via $\pmb{\Gamma_\lambda}$. 

Even when the state space of each process has different cardinality, it is still possible to establish a link via a local transformation. 

This connection also could be explored when the state space is time dependent, in the sense that the number of states change with time, that was used in \cite{caraj}, intuitively.
  
In addition, allows us to address a new problem from another known one, through a non-local modulation of the well known solution, following the expression \eqref{conlc}. We have not only shown that this is feasible to do through formal and constructive proof of existence of that $\pmb\lambda$, but also we indicated what is the right way to do it, should be across a linear and local (time-dependent) operation. 

Future research through a Lagrangian description (working process) may find novel applications of the present proposal. In this approach the role of this kind of transformation $\pmb{\lambda}$ will be a symmetry of the Lagrangian. 
A gauge theory of stochastic processes can be improved formally through a variational principle.

\vspace*{0.5cm}

\section*{Acknowledgments}

We thank our respective institutions, UNQ-IFIBA, UGR and CONICET and a special mention to Fernando Cornet. Recall also our anonymous readers and reviewers for their contribution to this work, including Federico G. Vega. Gabriel Lio, Micaela Moretton and Mar\'ia Clara Caruso have shared with us the labor as local coaches.

\vspace*{0.4cm}
\section*{Appendix}
\appendix*
\vspace*{0.5cm}
\setcounter{equation}{0}

\section*{\sc \textbf{A1} Some relevant properties of map  $\pmb{\Gamma_\lambda}$}

$\bullet$ \pmb{\sc \footnotesize COMPOSITE MAPS}\\

We composed two transformation $\pmb{\Gamma_{\lambda}}\circ\pmb{\Gamma_{\lambda}'}$ with
$\pmb{\lambda}$ and $\pmb{\lambda}'$ are non singular and then prove that 
\begin{equation}\label{composite}
\pmb{\Gamma_{\lambda}}\circ\pmb{\Gamma}_{\pmb{\lambda}'}(\pmb{\mathtt{Q}})=\pmb{\Gamma}_{\pmb{\lambda}\pmb{\lambda}'}(\pmb{\mathtt{Q}}).
\end{equation}
We calculate directly 

\begin{align*}
\pmb{\Gamma_{\lambda}}\circ\pmb{\Gamma}_{\pmb{\lambda}'}(\pmb{\mathtt{Q}})
& = \pmb{\Gamma_{\lambda}}(\pmb{\lambda}' \pmb{\mathtt{Q}} \pmb{\lambda}'^{-1} +d_t\pmb{\lambda}'\pmb{\lambda}'^{-1})\nonumber\\
& = \pmb{\lambda}[\pmb{\lambda}' \pmb{\mathtt{Q}} \pmb{\lambda}'^{-1} +d_t\pmb{\lambda}'\pmb{\lambda}'^{-1}]\pmb{\lambda}^{-1}
+d_t\pmb{\lambda}\pmb{\lambda}^{-1}\nonumber\\
& = \pmb{\lambda}\pmb{\lambda}' \pmb{\mathtt{Q}} (\pmb{\lambda\lambda}')^{-1}
+\pmb{\lambda}d_t\pmb{\lambda}'(\pmb{\lambda}\pmb{\lambda}')^{-1}
+d_t\pmb{\lambda}\pmb{\lambda}'(\pmb{\lambda}\pmb{\lambda}')^{-1}\nonumber\\
& = \pmb{\lambda}\pmb{\lambda}' \pmb{\mathtt{Q}} (\pmb{\lambda\lambda}')^{-1}
+d_t(\pmb{\lambda}\pmb{\lambda}')(\pmb{\lambda}\pmb{\lambda}')^{-1}\nonumber\\
& = \pmb{\Gamma}_{\pmb{\lambda}\pmb{\lambda}'}(\pmb{\mathtt{Q}}),\nonumber
\end{align*}
where in the second line the term $d_t\pmb{\lambda}'\pmb{\lambda}'^{-1}$ is written as $d_t\pmb{\lambda}'(\pmb{\lambda}\pmb{\lambda}^{-1})\pmb{\lambda}'^{-1}$. This complete the demostration that expression \eqref{composegamma} is satisfied.\\

$\bullet$ \pmb{\sc \footnotesize INVERSE MAP}\\

We calculate explicitly $\pmb{\Gamma}_{\pmb{\lambda}^{-1}}(\pmb{\mathtt{Q}})$ and then prove that 
\begin{equation}\label{invers}
\pmb{\Gamma}_{\pmb{\lambda}^{-1}}(\pmb{\mathtt{Q}})=\pmb{\Gamma}_{\pmb{\lambda}}^{-1}(\pmb{\mathtt{Q}}),
\end{equation} 
for all $\pmb{\mathtt{Q}}$. Let's calculate the left hand of \eqref{invers}
\begin{align*}
\pmb{\Gamma}_{\pmb{\lambda}^{-1}}(\pmb{\mathtt{Q}})
&=\pmb{\lambda}^{-1} \pmb{\mathtt{Q}} \pmb{\lambda} +d_t(\pmb{\lambda}^{-1})\pmb{\lambda}\nonumber\\
&=\pmb{\lambda}^{-1} \pmb{\mathtt{Q}} \pmb{\lambda} +d_t(\pmb{\lambda}^{-1})\pmb{\lambda}\nonumber\\
&=\pmb{\lambda}^{-1} \pmb{\mathtt{Q}} \pmb{\lambda} -\pmb{\lambda}^{-1}d_t\pmb{\lambda}.
\end{align*}

Finally we check directly that $\pmb{\Gamma}_{\pmb{\lambda}^{-1}}$ is equal to $\pmb{\Gamma}_{\pmb{\lambda}}^{-1}$ 
\begin{align*}
\pmb{\Gamma_{\lambda}}\circ \pmb{\Gamma}_{\pmb{\lambda}^{-1}}(\pmb{\mathtt{Q}})
&=\pmb{\Gamma_{\lambda}}(\pmb{\lambda}^{-1} \pmb{\mathtt{Q}} \pmb{\lambda} -\pmb{\lambda}^{-1}d_t\pmb{\lambda})\nonumber\\
&=\pmb{\lambda} (\pmb{\lambda}^{-1} \pmb{\mathtt{Q}} \pmb{\lambda} -\pmb{\lambda}^{-1}d_t\pmb{\lambda})\pmb{\lambda}^{-1}+  d_t\pmb{\lambda} \pmb{\lambda}^{-1}\nonumber\\
&=\pmb{\lambda} \pmb{\lambda}^{-1} \pmb{\mathtt{Q}} \pmb{\lambda} \pmb{\lambda}^{-1} -\pmb{\lambda}\pmb{\lambda}^{-1}d_t\pmb{\lambda}\pmb{\lambda}^{-1}+  d_t\pmb{\lambda} \pmb{\lambda}^{-1}\nonumber\\
&= \pmb{\mathtt{Q}} -d_t\pmb{\lambda}\pmb{\lambda}^{-1}+  d_t\pmb{\lambda} \pmb{\lambda}^{-1}\nonumber \\
&= \pmb{\mathtt{Q}},
\end{align*}
where we used $d_t(\pmb{\lambda}^{-1}\pmb{\lambda})=\pmb{0} \Longrightarrow d_t(\pmb{\lambda}^{-1})\pmb{\lambda}=-\pmb{\lambda}^{-1}d_t\pmb{\lambda}$. 
This complete the demonstration that \eqref{Gammainv} is true, i.e.  $\pmb{\Gamma}_{\pmb{\lambda}^{-1}}=\pmb{\Gamma}_{\pmb{\lambda}}^{-1}$. \\

$\bullet$ \pmb{\sc \footnotesize $\Gamma_\lambda$ AS AN EQUIVALENCE RELATION}\\

We say that the map $\pmb{\Gamma_\lambda}$ define an equivalence relation between the space of between of the vector space of matrices of the same dimension.
For a given two matrices $(\pmb{\mathtt{Q}},\pmb{\mathtt{Q}}')$ we can define a relation between them 
\begin{align}
\pmb{\mathtt{Q}}'\sim \pmb{\mathtt{Q}}\Longleftrightarrow \exists \pmb{\lambda}: \pmb{\mathtt{Q}}'=\pmb{\Gamma}_{\pmb{\lambda}}(\pmb{\mathtt{Q}}),
\end{align}
where $\pmb{\Gamma}_{\pmb{\lambda}}(\pmb{\mathtt{Q}}):=\pmb{\lambda}\pmb{\mathtt{Q}}\pmb{\lambda}^{-1}+d_t\pmb{\lambda}\pmb{\lambda}^{-1}$ and $\pmb{\lambda}$ is a non singular matrix.  This relation $\sim $ is an equivalence in the sense that  $\forall$ $	\pmb{\mathtt{Q}}, \pmb{\mathtt{Q}}', \pmb{\mathtt{Q}}''$ the following properties are true: 
\begin{align*}
(\pmb{R})\quad&\pmb{\mathtt{Q}}\sim \pmb{\mathtt{Q}}\quad (\mathit{reflexivity}) \\
(\pmb{S})\quad&\pmb{\mathtt{Q}}\sim \pmb{\mathtt{Q}}'\Rightarrow \pmb{\mathtt{Q}}'\sim \pmb{\mathtt{Q}}\quad(\mathit{symmetry})\\
(\pmb{T})\quad&\pmb{\mathtt{Q}}''\sim \pmb{\mathtt{Q}}'\land \pmb{\mathtt{Q}}'\sim \pmb{\mathtt{Q}}\Rightarrow\pmb{\mathtt{Q}}''\sim \pmb{\mathtt{Q}}\quad (\mathit{transitivity})
\end{align*}

The first assertion ($\pmb{R}$) is true from the identity matrix $\pmb{\lambda}=\textbf{1}$ and by definition $\pmb{\Gamma}_{\pmb{1}}(\pmb{\mathtt{Q}})=\pmb{\mathtt{Q}}$. The assertion ($\pmb{S}$) is also true from the existence of the inverse matrix $\pmb{\lambda}^{-1}$ and construct through \eqref{invers} the \textit{inverse} connection  $\pmb{\mathtt{Q}}'\sim \pmb{\mathtt{Q}}$. The last assertion ($\pmb{T}$) is true from the composed transformation of non singular matrices $\pmb{\lambda}=\pmb{\lambda}''\pmb{\lambda}'$ and \eqref{composite}, such that $\pmb{\mathtt{Q}}''=\pmb{\Gamma}_{\pmb{\lambda}''}(\pmb{\mathtt{Q}}')$ and $\pmb{\mathtt{Q}}'=\pmb{\Gamma}_{\pmb{\lambda}'}(\pmb{\mathtt{Q}})$, then $\pmb{\mathtt{Q}}''=\pmb{\Gamma}_{\pmb{\lambda}''}(\pmb{\Gamma}_{\pmb{\lambda}'}(\pmb{\mathtt{Q}}))=\pmb{\Gamma}_{\pmb{\lambda}''\pmb{\lambda}'}(\pmb{\mathtt{Q}})=\pmb{\Gamma}_{\pmb{\lambda}}(\pmb{\mathtt{Q}})$. Finally, we get to $\pmb{\mathtt{Q}}''\sim \pmb{\mathtt{Q}}$.

\section*{\sc \textbf{A2} Alternative expression for gauge transformation $\pmb{\lambda}$}

In the present work we said that in order to give an expression for the solution of \eqref{lambdaeq}, we see that the \textit{transitivity solution} is constructed from the composed transformation $\pmb{\lambda}=\pmb{\lambda}_1\pmb{\lambda}_2$ and  \eqref{composite}

\begin{equation}
\pmb{\Gamma}_{\pmb{\lambda}_1\pmb{\lambda}_2}(\pmb{\mathtt{Q}})=\pmb{\Gamma}_{\pmb{\lambda}_1}\circ \pmb{\Gamma}_{\pmb{\lambda}_2}(\pmb{\mathtt{Q}})=\pmb{\mathtt{Q}}',
\end{equation}
where $\pmb{\lambda}_1$ and $\pmb{\lambda}_2$ are solution of  \eqref{l1} and \eqref{l2}, respectively
\begin{align}\label{lambdacompp}
\pmb{\mathtt{Q}}'\sim \pmb{0}\Longleftrightarrow & d_t\pmb{\lambda}_1=\pmb{\mathtt{Q}}'\pmb{\lambda}_1,\nonumber\\
&\\
\pmb{0}\sim \pmb{\mathtt{Q}}\Longleftrightarrow & d_t\pmb{\lambda}_2=-\pmb{\lambda}_2\pmb{\mathtt{Q}}.\nonumber
\end{align}

We express the solutions of \eqref{lambdacompp} as a formal iterative solution 
\begin{align}\label{iterativelambda}
\pmb{\lambda}_1(t)=&\bigg[1+\int_0^t \pmb{\mathtt{Q}}(t_1) dt_1\nonumber\\
&+\int_0^t\int_0^{t_1}\pmb{\mathtt{Q}}(t_1)\pmb{\mathtt{Q}}(t_2) dt_1 dt_2 +\cdots\bigg] \pmb{\lambda}_1(0),\nonumber \\
&\\
\pmb{\lambda}_2(t)=&\pmb{\lambda}_2(0)\bigg[1-\int_0^t \pmb{\mathtt{Q}}'(t_1) dt_1\nonumber\\
&+\int_0^t\int_0^{t_1}\pmb{\mathtt{Q}}'(t_1)\pmb{\mathtt{Q}}'(t_2) dt_1 dt_2 +\cdots\bigg].\nonumber
\end{align}

We obtain a general expression of the iterative solution \eqref{iterativelambda} through a Magnus series \cite{Magnus}
\begin{align}
\pmb{\lambda}_1(t)=& \sum_{n\in \mathbb{N}}\frac{1}{n!}\;\pmb{\Lambda}_n(t),\nonumber\\
&\\
\pmb{\lambda}_2(t)=& \sum_{n\in \mathbb{N}}\frac{(-1)^{n}}{n!}\pmb{\Lambda}'_n(t),\nonumber
\end{align}
where each $\pmb{\Lambda}_n(t)$ and $\pmb{\Lambda}'_n(t)$ are given by
\begin{align}
\pmb{\Lambda}_n(t)&=\int_0^t \pmb{\mathtt{Q}}(t_1)dt_1\int_0^{t_1} \pmb{\mathtt{Q}}(t_2)dt_2\cdots \int_0^{t_{n-1}} \pmb{\mathtt{Q}}(t_{n-1})dt_n,\nonumber\\
&\\
\pmb{\Lambda}'_n(t)&=\int_0^t \pmb{\mathtt{Q}}'(t_1)dt_1\int_0^{t_1} \pmb{\mathtt{Q}}'(t_2)dt_2\cdots \int_0^{t_{n-1}} \pmb{\mathtt{Q}}'(t_{n-1})dt_n.\nonumber
\end{align}

\end{document}